\documentclass[12pt]{article}
\usepackage{graphicx}
\usepackage[small,bf]{caption}
\usepackage{hyperref}

\catcode`\@=11 \long\def\@makefntext#1{ \protect\noindent \hbox to
3.2pt {\hskip-.9pt
$^{{\ninerm\@thefnmark}}$\hfil}#1\hfill}                

\def\@makefnmark{\hbox to 0pt{$^{\@thefnmark}$\hss}}  

\def\ps@myheadings{\let\@mkboth\@gobbletwo
\def\@oddhead{\hbox{}
\rightmark\hfil\ninerm\thepage}
\def\@oddfoot{}\def\@evenhead{\ninerm\thepage\hfil
\leftmark\hbox{}}\def\@evenfoot{}
\def\sectionmark##1{}\def\subsectionmark##1{}}

\renewcommand{\thefootnote}{\fnsymbol{footnote}}

\newcounter{sectionc}\newcounter{subsectionc}\newcounter{subsubsectionc}
\renewcommand{\section}[1] {\vspace*{0.6cm}\addtocounter{sectionc}{1}
\setcounter{subsectionc}{0}\setcounter{subsubsectionc}{0}\noindent
        {\normalsize\bf\thesectionc. #1}\par\vspace*{0.4cm}}
\renewcommand{\subsection}[1] {\vspace*{0.6cm}\addtocounter{subsectionc}{1}
        \setcounter{subsubsectionc}{0}\noindent
        {\normalsize\it\thesectionc.\thesubsectionc. #1}\par\vspace*{0.4cm}}
\renewcommand{\subsubsection}[1]
{\vspace*{0.6cm}\addtocounter{subsubsectionc}{1}
        \noindent
{\normalsize\rm\thesectionc.\thesubsectionc.\thesubsubsectionc.
        #1}\par\vspace*{0.4cm}}

\renewenvironment{thebibliography}[1]
        {\begin{list}{\arabic{enumi}.}
        {\usecounter{enumi}\setlength{\parsep}{0pt}
\setlength{\leftmargin 0.52cm}{\rightmargin 0pt}
         \setlength{\itemsep}{0pt} \settowidth
        {\labelwidth}{#1.}\sloppy}}{\end{list}}

\newcounter{itemlistc}
\newcounter{romanlistc}
\newcounter{alphlistc}
\newcounter{arabiclistc}

\def\@citex[#1]#2{\if@filesw\immediate\write\@auxout
        {\string\citation{#2}}\fi
\def\@citea{}\@cite{\@for\@citeb:=#2\do
        {\@citea\def\@citea{,}\@ifundefined
        {b@\@citeb}{{\bf ?}\@warning
        {Citation `\@citeb' on page \thepage \space undefined}}
        {\csname b@\@citeb\endcsname}}}{#1}}

\newif\if@cghi
\def\cite{\@cghitrue\@ifnextchar [{\@tempswatrue
        \@citex}{\@tempswafalse\@citex[]}}
\def\citelow{\@cghifalse\@ifnextchar [{\@tempswatrue
        \@citex}{\@tempswafalse\@citex[]}}
\def\@cite#1#2{{$\null^{#1}$\if@tempswa\typeout
        {IJCGA warning: optional citation argument
        ignored: `#2'} \fi}}

 1 
scaled\magstep 1  1
 
scaled\magstephalf

 \font\ninerm=cmr9

\newcommand{\beq}{\begin{equation}}
\newcommand{\eeq}{\end{equation}}

\newcommand\mathC{\mkern1mu\raise2.2pt\hbox{$\scriptscriptstyle|$}
                {\mkern-7mu\rm C}}

\begin{document}

\hfill \vspace*{1cm}

\centerline{\normalsize\bf ON THE EXISTENCE OF EXOTIC MATTER IN CLASSICAL}
\centerline{\normalsize\bf NEWTONIAN MECHANICS}

\vspace*{0.6cm} \centerline{\footnotesize SABBIR~A.~RAHMAN}
\vspace*{0.2cm} 
\centerline{\footnotesize\em Domestic Analysis Department, Strategy \& Market Analysis Division, }
\centerline{\footnotesize\em Saudi Aramco, Dhahran 31311, Eastern Province, Kingdom of Saudi Arabia}
\vspace*{0.1cm} \centerline{\footnotesize \em E-mail:
sarahman@alum.mit.edu} \vspace*{0.9cm}

{\centering{\begin{minipage}{12.2truecm}\footnotesize\baselineskip=12pt
\noindent\centerline{\footnotesize ABSTRACT} \vspace*{0.1cm}
\parindent=0pt

According to Newton's law of gravitation the force between two particles depends upon
their inertial, as well as their active and passive gravitational masses. For ordinary matter
all three of these are equal and positive. We consider here the more general case
where these quantities are equal in magnitude for a given particle but can differ in sign. The
resulting set of possible interactions allows each particle type to be assigned
to one of precisely four different classes, and the results of N-body simulations show that
the corresponding dynamics can give rise to a fairly rich spectrum of possible outcomes, some of which
are familiar from nature at various scales. Total energy and momentum are conserved by all of these
interactions if the definitions of momentum and kinetic and potential energy are suitably
generalised.

\end{minipage}\par}}
\vspace*{0.6cm}

\normalsize\baselineskip=15pt \setcounter{footnote}{0}
\renewcommand{\thefootnote}{\alph{footnote}}    

\section{Introduction}

The nature of mass and how the conception of mass has evolved over time has been discussed in some detail in two classic works by Jammer\cite{Jammer61,Jammer99}.
%
For the matter that we observe on a day-to-day basis, accurate experiments have been carried out which establish the equality, for any given body, of its inertial, and active and passive gravitational masses\cite{Bondi57}. However one cannot rule out the possibility of the existence of exotic types of matter for which this equality does not hold, and indeed physicists at various times have speculated about the existence of particles of negative inertial or gravitational mass\cite{Bondi57}\textsuperscript{-}\cite{Bonnor89}, or of mass which antigravitates.

The current general concensus tends to be that both Einstein's equivalence principle and Newton's third law both hold and that there are no antigravitating particles, so that the inertial, active and passive masses are equal, and that mass is positive for all particles in nature. This opinion was discussed in detail by Nieto and Goldman\cite{Nieto91} and their critique remains a key reference in this regard. The possibility of the existence of negative mass or antigravitation has been revived recently by a number of authors\cite{Chardin97}\textsuperscript{-}\cite{Chardin18}, and matter with exotic properties are occasionally introduced into hypothetical cosmological models\cite{Arun17} but  the actual existence of such exotic types of matter in nature remains rather speculative.

Nevertheless, it has yet to be proven experimentally for example that antimatter does not antigravitate, because of the inherent difficulty in isolating and carrying out gravitational measurements on antimatter. It is only recently that it has been possible to isolate significant quantities of antihydrogen, and a number of experiments are underway\cite{Kellerbauer08}\textsuperscript{-}\cite{Bertsche18} to determine whether the principle of equivalence holds for antimatter. The sensitivity of these experiments is not yet sufficiently high to answer this question but it is hoped that a definitive answer will be forthcoming in due course.

The later work by Jammer\cite{Jammer99} analysed specific examples of what would happen if for example active and passive gravitational mass happens to have negative sign for some particles, and described the kind of non-intuitive dynamics that would result, even if energy and momentum happened to be conserved. This of course would maximally violate the weak equivalence principle of general relativity. Other combinations of signs for the various types of masses would also appear to violate conservation of energy or momentum as these quantities are usually defined, or would violate Newton's third law of motion. Nevertheless, due perhaps to the number of possible particle types and the corresponding interactions that would need to be considered, and given that those interactions that have been considered appear to have highly non-intuitive or unfamiliar properties, no attempt appears to have been made thus far to carry out an exhaustive analysis of the various possibilities.

In this paper, we carry out a detailed analysis of the case wherein the inertial, active and passive masses of each particle are restricted to have the same magnitude but can each independently have a positive or negative sign. Allowing the mass coefficient itself also to have either positive or negative sign gives rise to 16 different possible particle types and we establish the complete set of possible interactions between them. We show that it is possible to come up with a general definition of momentum, and kinetic and gravitational potential energy such that total momentum and energy are conserved by all of these interactions, and state the appropriate generalisation of Newton's third law\footnote{A parallel work recently published by Chardin, Manfredi {\it et al.}\cite{Manfredi18} covers similar ground to this paper, but with a different perspective focusing on large scale structure and cosmology, whereas our focus here is mainly at the level of the substructure of elementary particles. They consider an additional four classes of exotic particle not considered here whose interactions with ordinary matter cannot be accounted for merely by changing the signs of the inertial, active and passive masses in Newton's law, and whose interactions also do not satisfy the principle of equivalence. While this increases the space of possible configurations, it is not immediately clear that these can be made compatible with the usual conservation laws, or with general relativity.}.

We also carry out N-body simulations of the many particle case where one or more types of particle are present and survey the behaviour of the resulting dynamics and the kinds of large scale structure that can arise. Although the interactions between individual pairs of particles may seem non-intuitive based upon our day-to-day observations of ordinary matter, we find that many of the resulting many-body configurations are actually quite familiar to us from observations of nature varying in scale from condensed matter systems to the large scale structure of the universe. This is quite suggestive of the possibility that exotic matter of the types considered may in reality have an important role to play at both the largest and smallest scales.

\section{A review of the concept of mass in Newtonian physics}

According to Newton's law of gravity, the force on a particle of mass $m_1$ at position ${\bf r}_1$ due to a particle of mass $m_2$ at position ${\bf r}_2$ is given by the well-known force law,

\begin{equation}
{\bf F}_{12} = m_1 {\bf a}_{12} =\frac{G m_1 m_2}{r_{12}^2}\,{\hat{\bf r}}_{12}\,,
\end{equation} 
where ${\bf F}_{12}$ is the gravitational force acting on particle 1, ${\bf a}_{12}$ is its resulting acceleration, $G$ is Newton's gravitational constant, ${\hat{\bf r}}_{12}$ is the unit vector in the direction ${\bf r}_2-{\bf r}_1$ and $r_{12}=|{\bf r}_2-{\bf r}_1|$.

As pointed out by Bondi\cite{Bondi57}, there are actually three different types of mass that enter into this equation, namely, (i) the active gravitational mass of particle 2 that is the source of the gravitational field, (ii) the passive gravitational mass of particle 1 that is acted upon by the field due to particle 2, and (iii) the inertial mass of particle 1 that determines the acceleration of particle 1 in response to the gravitational force that it experiences.

All three are usually assumed to be equal and positive. However experimental confirmation of this has thus far only been carried out on ordinary matter, and it may not be true for antimatter or other types of exotic matter should they exist. Exotic matter can in general have other combinations of the three mass types and we will investigate these possibilities further in what follows.

\subsection{Active, passive, and inertial mass}

We explore here the various types of interactions and particle configurations that are possible if, for a given particle, the three mass types have the same magnitude, but can each separately have positive or negative sign, i.e. inertial mass = $Im$, active gravitational mass = $Am$ and passive gravitational mass = $Pm$, where $I$, $A$ and $P$ are signs multiplying an underlying `core' mass $m$ identified with the particle. The mass $m$ could itself also be positive or negative, and so we allow for an additional sign factor $S$, such that $m=S|m|$. This then gives rise to sixteen possible particle types\footnote{Jammer\cite{Jammer99} only considers eight particle types as $m$ is taken to be positive throughout.} that can be denoted by the quadruple $(I, A, P, S)$, where each element of the tuple is either `$+$' or `$-$' and represents the sign of the inertial mass, the active gravitational and passive gravitational masses, and the core mass, respectively.

Given two particles of type $(I_1, A_1, P_1, S_1)$ and  $(I_2, A_2, P_2, S_2)$ respectively, the force on particle 1 due to particle 2 is given by,
\begin{equation}
{\bf F}_{12} \equiv I_1 m_1 {\bf a}_{12} =\frac{G P_1 m_1 A_2 S_2 |m_2|}{r_{12}^2}\,{\hat{\bf r}}_{12}\,.\label{eqn:forcedef}
\end{equation} 

This means that particle $1$ will be attracted to particle $2$ if the product $I_1 P_1 A_2 S_2$ is positive, and repelled by it if the product is negative. Note that the interactions of particles of type $(I, A, P, S)$ are indistinguishable from those of particles of type $(-I, A, -P, S)$ or of type $(I, -A, P, -S)$, and hence also of type $(-I, -A, -P, -S)$, so that in terms of interparticle interactions there are really only four classes of particle that need to be considered. We label these as classes {\bf A}, {\bf B}, {\bf C} and {\bf D} for convenience, with ordinary matter of type $(++++)$ being assigned to class {\bf A}. These are precisely the particle classes originally considered by Bondi\cite{Bondi57}, though in that work the scope was restricted to only those particle types with equal active and passive gravitational mass ($A=P$), and positive core mass ($S=+1$).  Manfredi {\it et al.}\cite{Manfredi18} also consider the same four classes\footnote{Note however that our classes {\bf B}, {\bf C} and {\bf D} correspond to their cases {\bf C}, {\bf D} and {\bf B} respectively.}, but restrict their attention to the interactions of each class of exotic matter with ordinary matter, and do not consider the possibility of there being more than one type of exotic matter present.

An important observation to be made here is that all of the particles belonging to classes {\bf B}, {\bf C} and {\bf D} - and even three of the four particle types belonging to class {\bf A} - have {\it some} kind of negative mass, and so one needs to be very careful to distinguish between the various types when referring to, or making general statements about, particles of `negative mass'. Unfortunately such care has not always been observed in the literature in the past, leading to some general confusion on the matter.

The form of Eqn.(\ref{eqn:forcedef}) requires that Newton's third law, stating that for every action there is an equal and opposite reaction, be generalised to,
\begin{equation}
P_1A_2{\bf F}_{12} = -P_2A_1\,{\bf F}_{21}\,,\label{eqn:thirdlaw}
\end{equation}
and so it is natural to {\it define} the `action of particle 2 on particle 1' as the expression on the left hand side of this equation.

Table \ref{tab:table1} lists the particle types that belong to each class, and also shows whether a particle belonging to any particular class is attracted to, or repelled by, a particle of any other class. The various interactions between particle classes are represented more visually in Table \ref{tab:table2}.

\bgroup
\def\arraystretch{1.5}
\begin{table}
  \begin{center}
    \begin{tabular}{|l|c|c|c|c|c|c|c|c} 
      \cline{1-5}\cline{7-8}
      \textbf{Particle class} $(IAPS)$ - Interactions \& Potential& \textbf{A} & \textbf{B} & \textbf{C} & \textbf{D} && $IP$ & $AS$ \\
      \cline{1-5}\cline{7-8}
      \textbf{A} $=\{(++++),(-+-+),(+-+-),(----)\}$ & $+$ & $-$ & $+$ & $-$ && $+$ & $+$\\
      \textbf{B} $=\{(+-++),(---+),(+++-),(-+--)\}$ & $+$ & $-$ & $+$ & $-$ && $+$ & $-$\\
      \textbf{C} $=\{(++-+),(-+++),(+---),(--+-)\}$ & $-$ & $+$ & $-$ & $+$ && $-$ & $+$\\
      \textbf{D} $=\{(+--+),(--++),(++--),(-++-)\}$ & $-$ & $+$ & $-$ & $+$ && $-$ & $-$\\
      \cline{1-5}\cline{7-8}
    \end{tabular}
    \caption{Interactions between particle types arranged by class. Particles in each row are either attracted to ($+$), or repelled by ($-$), particles in each column.\label{tab:table1}}
  \end{center}
\end{table}
\egroup

\bgroup
\def\arraystretch{1.5}
\begin{table}
  \begin{center}
\begin{tabular}{|c|c|c|c|}
\hline
\quad\textbf{A}$\rightarrow$\quad$\leftarrow$\textbf{A}\quad&
\qquad\textbf{B}$\rightarrow$\quad\quad\textbf{A}$\rightarrow$&
$\leftarrow$\textbf{C}\qquad$\leftarrow$\textbf{A}\quad&
\quad$\leftarrow$\textbf{D}\quad\qquad\textbf{A}$\rightarrow$ \\

$\leftarrow$\textbf{A}\qquad$\leftarrow$\textbf{B}\quad&
\quad$\leftarrow$\textbf{B}\quad\qquad\textbf{B}$\rightarrow$&
\quad\textbf{C}$\rightarrow$\quad$\leftarrow$\textbf{B}\quad&
\qquad\textbf{D}$\rightarrow$\quad\quad\textbf{B}$\rightarrow$ \\

\qquad\textbf{A}$\rightarrow$\quad\quad\textbf{C}$\rightarrow$&
\quad\textbf{B}$\rightarrow$\quad$\leftarrow$\textbf{C}\quad&
\quad$\leftarrow$\textbf{C}\quad\qquad\textbf{C}$\rightarrow$&
$\leftarrow$\textbf{D}\qquad$\leftarrow$\textbf{C}\quad \\

\quad$\leftarrow$\textbf{A}\quad\qquad\textbf{D}$\rightarrow$&
$\leftarrow$\textbf{B}\qquad$\leftarrow$\textbf{D}\quad&
\qquad\textbf{C}$\rightarrow$\quad\quad\textbf{D}$\rightarrow$&
\quad\textbf{D}$\rightarrow$\quad$\leftarrow$\textbf{D}\quad \\
\hline
    \end{tabular}
    \caption{Direction of interparticle forces for pairs of particles in each class.\label{tab:table2}}
  \end{center}
\end{table}
\egroup

It can be seen by inspection of the interparticle forces in Table \ref{tab:table2} that for momentum to be conserved, particles of class {\bf D} must have the same sign of momentum as those of class {\bf A}, while the momentum of particles of class {\bf B} and {\bf C} must have the opposite sign. This will be the case if a particle of type $(I, A, P, S)$ with mass $m$ and velocity ${\bf v}$ has momentum $IAPm{\bf v}$. Similarly, by considering those interactions in which pairs of particles initially at rest accelerate together in the same direction (so that there is no change in the gravitational potential energy), energy conservation requires that the particle has kinetic energy $\frac{1}{2}IAPmv^2$,
\begin{equation}
E = {\textstyle\frac{1}{2}}IAPmv^2\,, \qquad {\bf p} = IAPm{\bf v}\,.\label{eqn:energymom}
\end{equation}

In order to deduce the form of the gravitational potential energy of one particle due to the gravitational field of another, one considers those interactions where two particles initially at rest are accelerated in opposite directions. In these cases there is always an increase in kinetic energy, and this must be coupled with a corresponding decrease in gravitational potential energy. Because particles sharing the same values for the products $IP$ and $AS$ belong to the same class and share the same interactions, the gravitational potential energy of particle 1 due to particle 2 must take the form,

\begin{equation}
U_{12} = -\frac{G(I_1P_1)^a(A_1S_1)^b(I_2P_2)^c(A_2S_2)^d|m_1||m_2|}{r_{12}}\,,
\end{equation}
where each of the exponents $a$, $b$, $c$ and $d$ can either be 0 or 1. The initial minus sign is necessary to ensure that the gravitational potential energy decreases when ordinary class {\bf A} particles that are initially at rest attract one another.

Similar consideration of the other interactions places constraints on various combinations of the exponents, and the only solutions that are consistent with the interactions turn out to be $(b=d=1, a=c=0)$ or $(a=c=1, b=d=0)$.

Now, the gravitational potential of a particle should depend upon its active mass, so the second of these two sets of solutions must be the correct one, implying that\footnote{The presence of only the magnitude of the mass $|m_1|$ here and also in Eqn.(\ref{eqn:gravpot}) justifies the separate inclusion of the sign factor $S$ associated with the core mass that has been omitted by others.},
\begin{equation}
 U_{12} = -\frac{G I_1P_1A_2S_2|m_1||m_2|}{r_{12}}\,.\label{eqn:pot12}
\end{equation}

 The sign of the gravitational potential of particle 1 in the field of particle 2 is therefore $-I_1P_1A_2S_2$ and thus the signs in Table \ref{tab:table1} indicating attraction or repulsion serve equally to indicate whether energy is required ($+$), or released ($-$), when separating a particle in the corresponding row from a particle in the corresponding column.

Thus the gravitational potential energy increases when two particles of class {\bf A} or of class {\bf D} are pulled apart, or when a particle of class {\bf B} is separated from a particle of class {\bf C}. Conversely, the potential energy decreases when two particles of class {\bf B} or of class {\bf C} are pushed apart, or when a particle of class {\bf A} is separated from a particle of class {\bf D}. Finally, because of the antisymmetry, there is no net change in gravitational potential energy if a particle of either class {\bf A} or {\bf D} is displaced relative to a particle of class {\bf B} or {\bf C}. 

It can be inferred from the second expression in (\ref{eqn:pot12}) that the spherically symmetric gravitational potential due to a particle of type $(I, A, P, S)$ and mass $m$ takes the form,

\begin{equation}
\Phi(r) = -\frac{GAm}{r}\,,
\end{equation}
where $Am$ is its active mass, which agrees with the usual definition. The gravitational potential energy of a particle of type $(I, A, P, S)$ and mass $m$ in a field $\Phi({\bf r})$ is then given by,
\begin{equation}
U({\bf r}) = IPSm\Phi({\bf r})\,,\label{eqn:gravpot}
\end{equation} 
where the nontrivial sign factor is required to ensure consistency. The force on the particle is given by,
\begin{equation}
{\bf F}({\bf r}) = -Pm\nabla\Phi({\bf r})\,,\label{eqn:force}
\end{equation} 
where $Pm$ is its passive mass, and again this agrees with the usual definition.

These results show that the classical concepts of momentum and kinetic and gravitational potential energy can be suitably generalised to the case of exotic particles of the types that we consider here without violating any of the usual laws or conservation principles of classical Newtonian mechanics. This should address any objections to the possible existence of such exotic particles in Newtonian mechanics on these grounds alone.

\begin{figure}
\begin{center}
\includegraphics[scale=0.62]{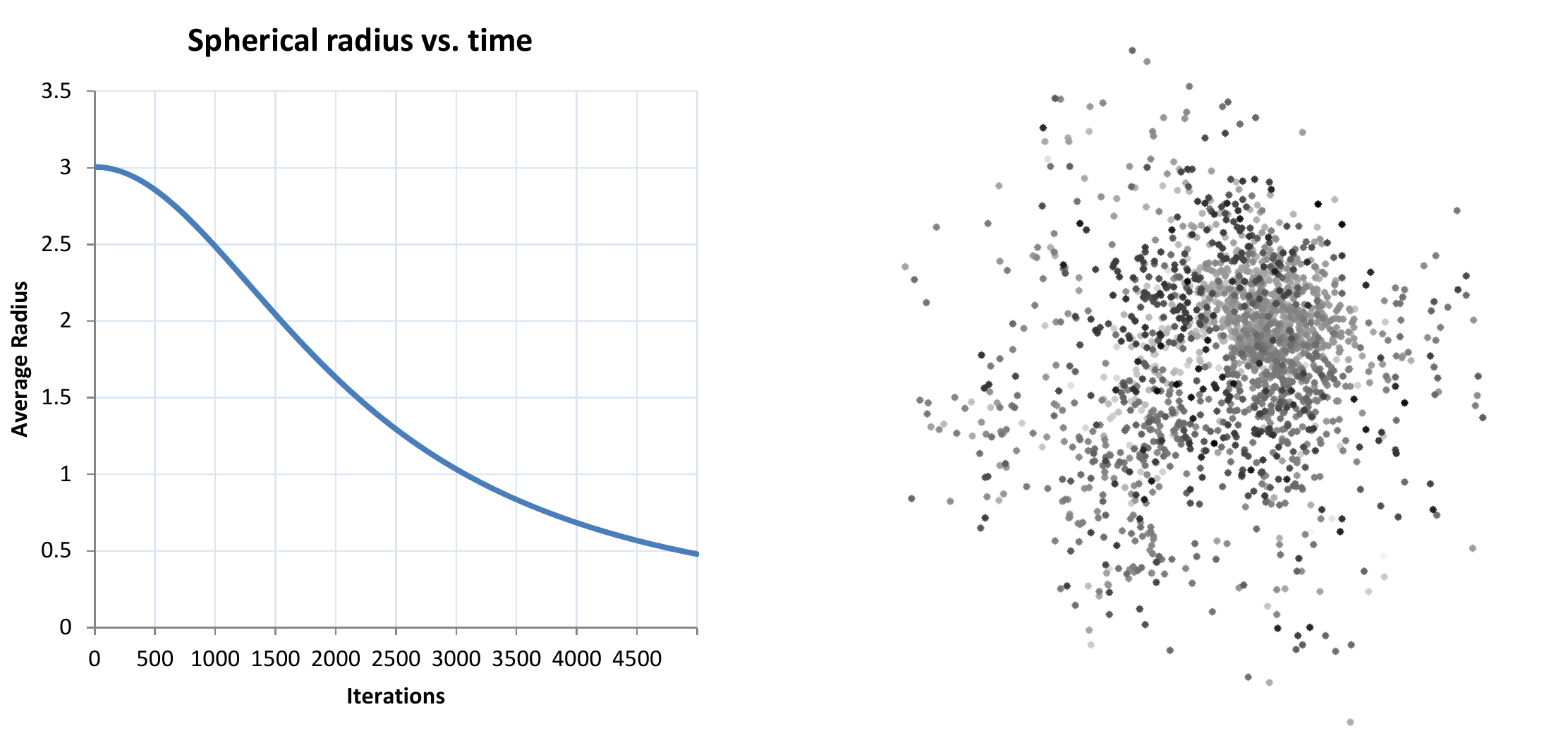}
\caption{When only particles of class {\bf A} or of class {\bf D} are present, matter will tend to clump into fractal-like clusters around regions of higher mass density, leaving behind increasingly large voids of lower mass density, so that that any inhomogeneities in density will increase over time. In an open universe, the mutual attraction between the particles leads them to contract towards the centre of mass.\label{fig:nbodyA}}
\vspace{10mm}
\includegraphics[scale=0.62]{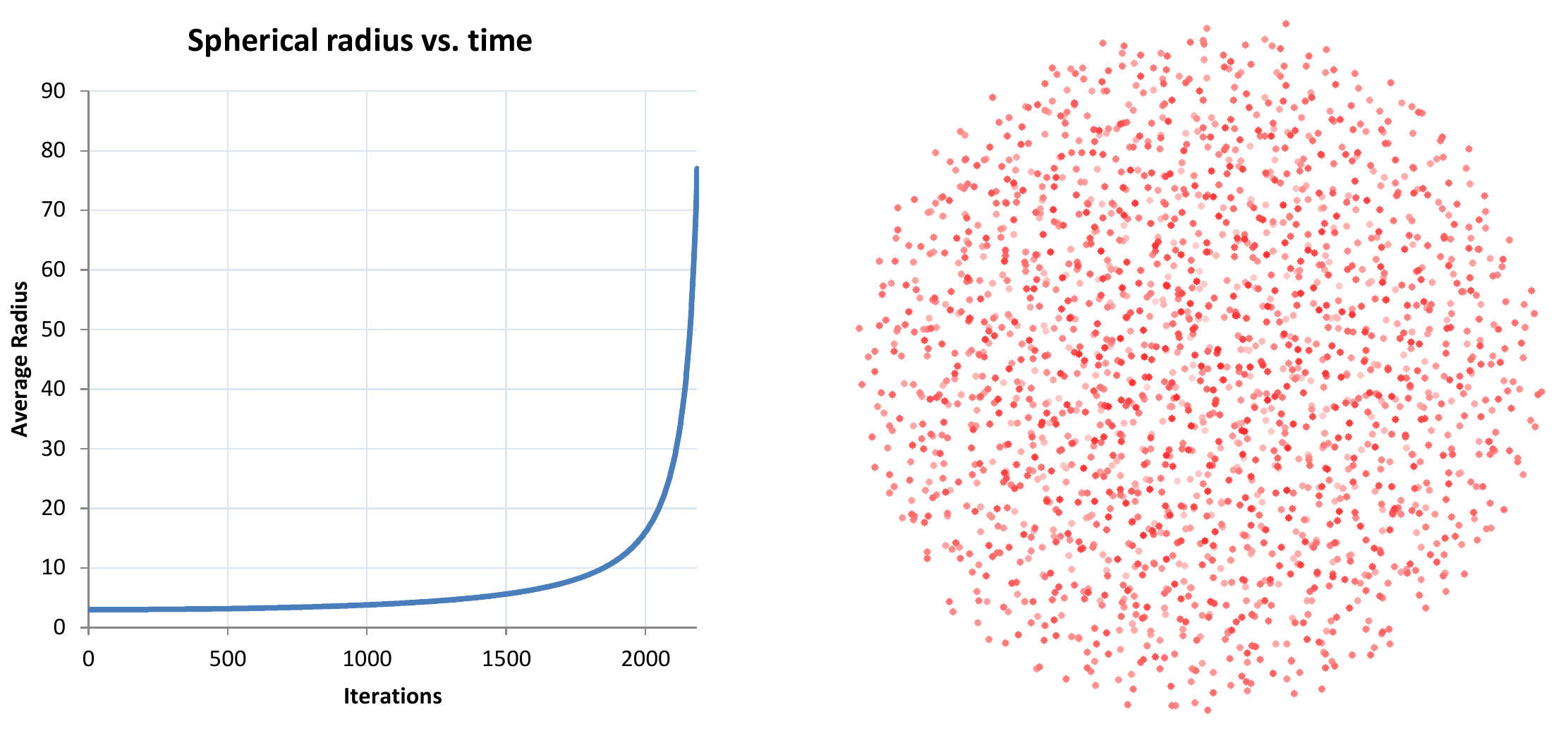}
\caption{When only particles of class {\bf B} or of class {\bf C} are present, their mutual repulsion will cause them to expand outwards and fill space with the rate of expansion accelerating over time. There will be a stronger mutual repulsion between particles in regions of higher density so that the overall effect will be to smooth out any non-uniformities over time.\label{fig:nbodyB}}
\end{center}
\end{figure}


\begin{figure}
\begin{center}
\includegraphics[scale=0.62]{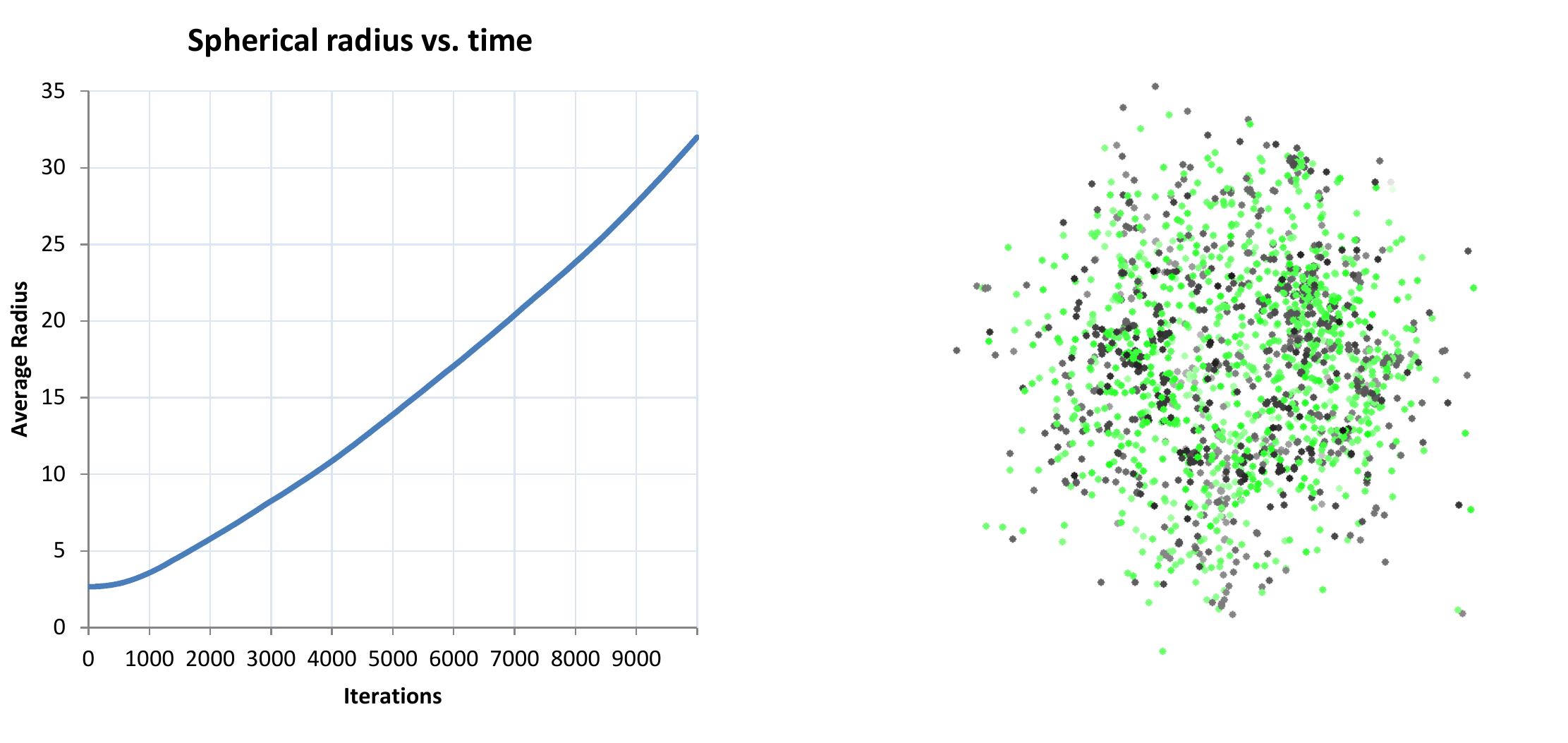}
\caption{When equal numbers of particles of class {\bf A} and {\bf B} are present, they gradually condense to form fractal-like clusters containing mostly particles of class {\bf A}, albeit somewhat more slowly than in the case when only particles of class {\bf A} were present. The attraction of particles of class {\bf B} to those of class {\bf A} is counteracted by their tendency to spread out and fill space, and consequently the formation of dense structures is accompanied by an overall gentle expansion. The resulting evolution is therefore not too dissimilar to that observed in nature where complex structures are being formed within an expanding universe. The same distributions will be observed if only particles of class {\bf C} and {\bf D} are present.\label{fig:nbodyAB}
}
\vspace{10mm}
\includegraphics[scale=0.62]{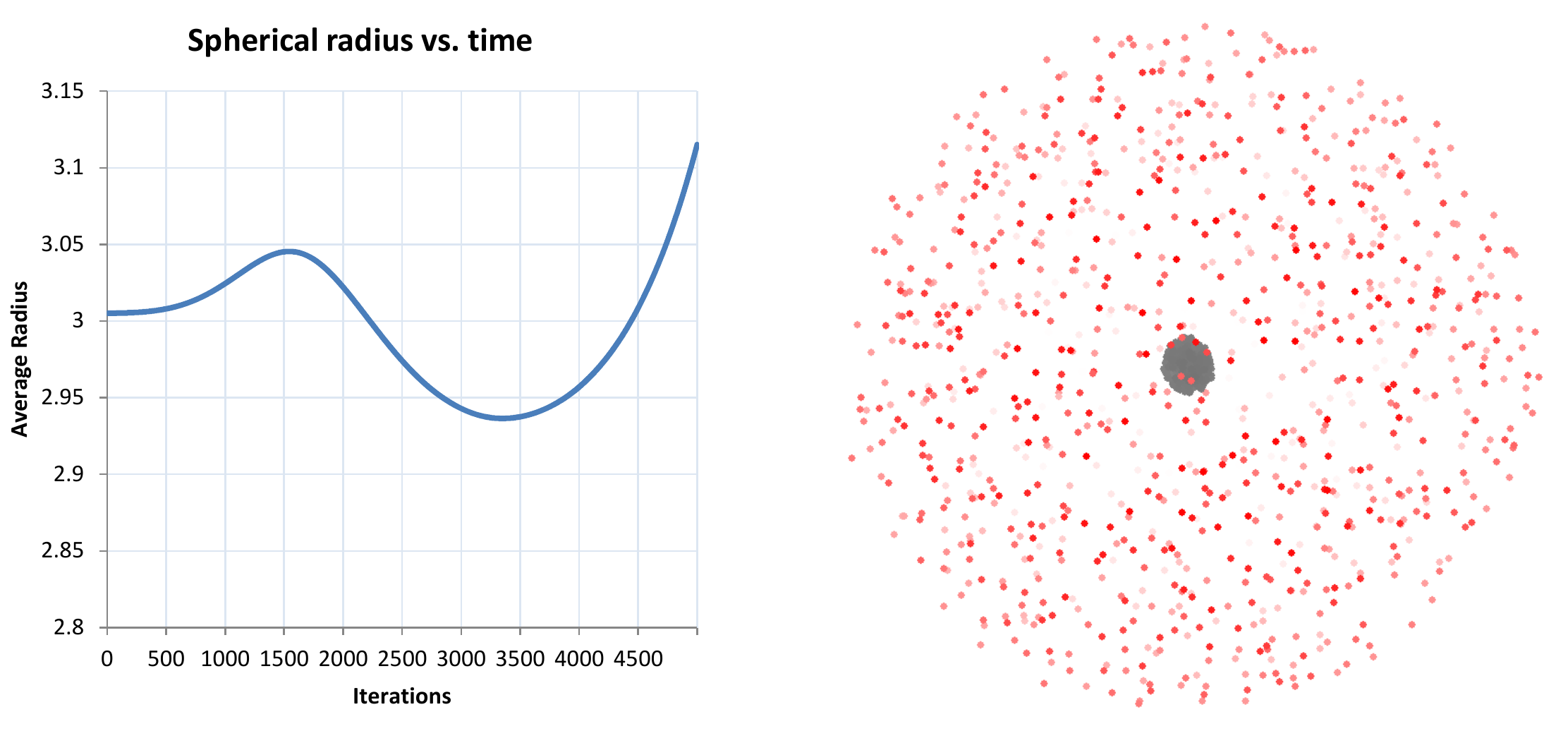}
\caption{When equal numbers of particles of class {\bf A} and {\bf C} are present, the particles of class {\bf C} will tend to spread outwards to fill space homogeneously while the particles of class {\bf A} will tend to form fractal-like localised clusters of increasing density. There is an initial period of slight expansion as the particles juggle around to make space for the rapid simultaneous collapse and expansion that follows. The same behaviour will be observed if only particles of class {\bf B} and {\bf D} are present.\label{fig:nbodyAC}}
\end{center}
\end{figure}

\begin{figure}
\begin{center}
\includegraphics[scale=0.62]{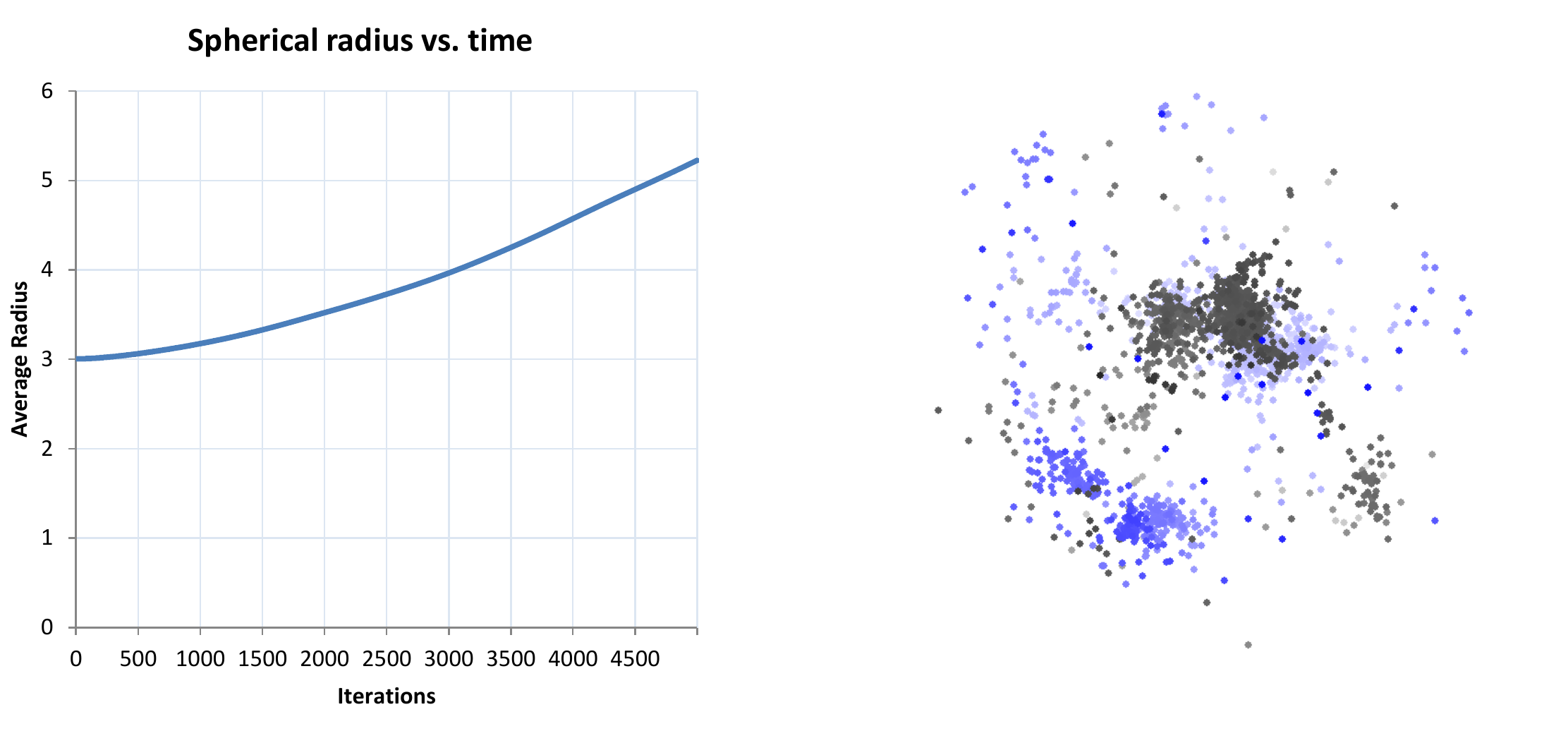}
\caption{For a mixture of particles of class {\bf A} and {\bf D}, particles of the same class attract each other, while particles of different classes mutually repel. In this case, space will gradually become partitioned into  (a perhaps fractal-like hierarchy of) domains of matter and antimatter (of negative gravitational mass), whose mutual repulsion will lead to an overall expansion of the universe. This configuration corresponds to the Janus cosmological model of Petit and d'Agostini.\label{fig:nbodyAD}}
\vspace{10mm}
\includegraphics[scale=0.62]{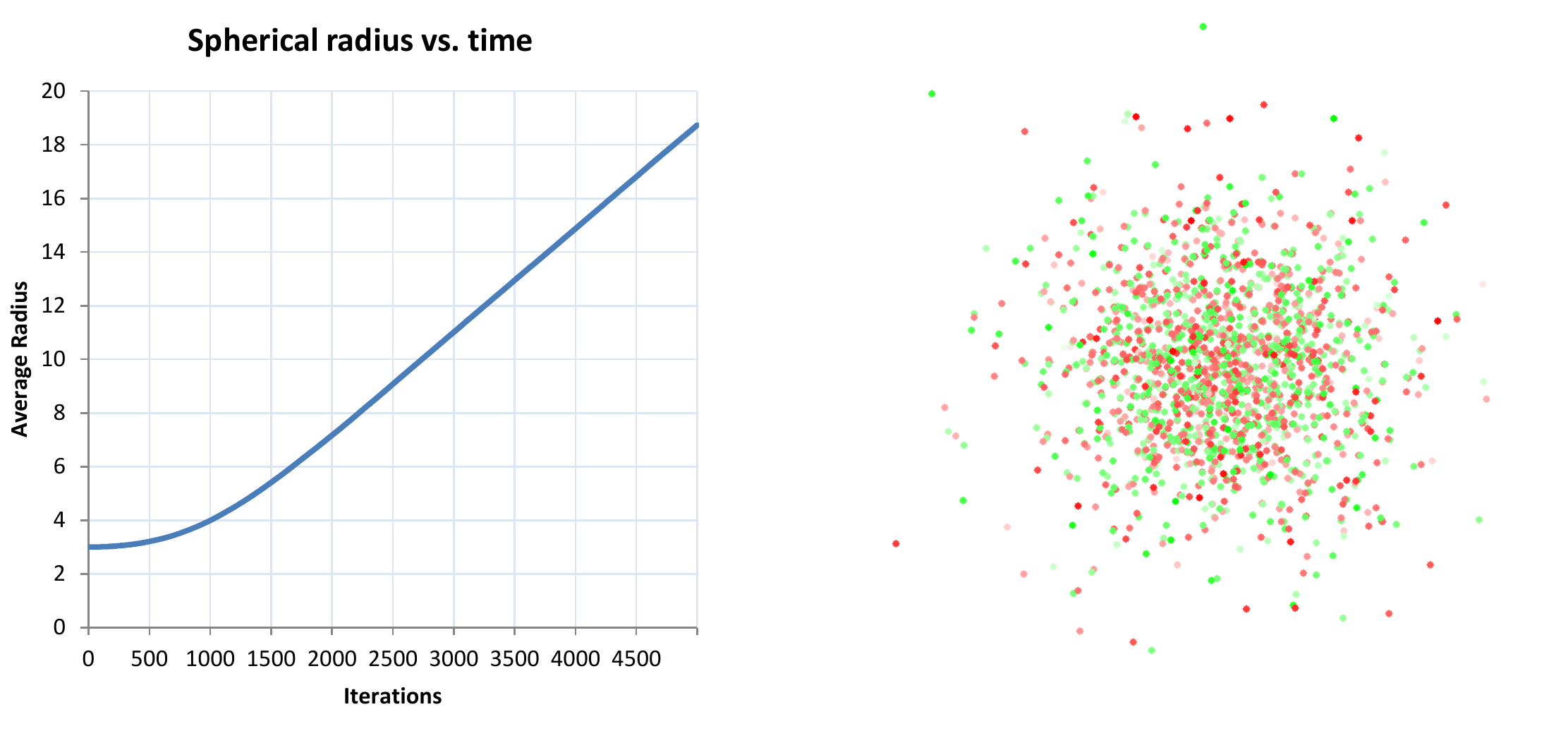}
\caption{When equal numbers of particles of class {\bf B} and {\bf C} are present, their interactions are akin to the electrostatic interactions between charges and the particles tend to intermix in a complex manner. There will be a tendency to form compact structures, but particles in regions of localised unequal charge density that inevitably form will tend to mutually repel resulting in an overall outward expansion.
\label{fig:nbodyBC}
}
\end{center}
\end{figure}

\begin{figure}
\begin{center}
\includegraphics[scale=0.62]{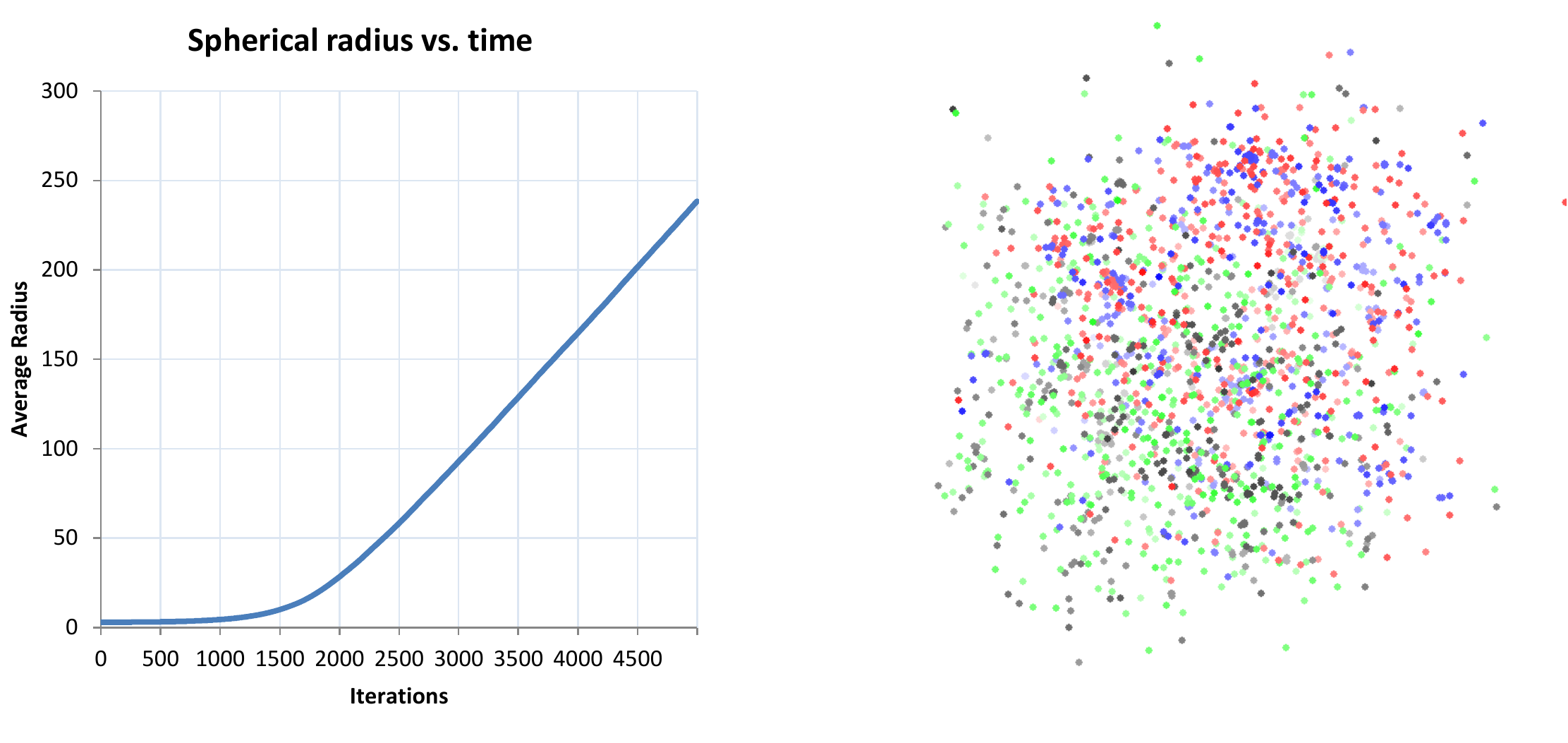}
\caption{When equal numbers of particles of class {\bf A}, {\bf B}, {\bf C} and {\bf D} are present, they form mutually repulsive domains with each domain consisting either of particles only of class {\bf A} and {\bf B}, or of particles of only class {\bf C} and {\bf D}. The resulting evolution will include a combination of the dynamics observed in Figures \ref{fig:nbodyAB} and \ref{fig:nbodyAD}. \label{fig:nbodyABCD}
}
\end{center}
\end{figure}

\subsection{The many-particle case}

The other common objection that is raised against the existence of such exotic particles is the apparently nonintuitive nature of their interactions. For example, if we consider the case of a particle of class {\bf A} and a particle of class {\bf B} initially at rest, these will accelerate off towards infinity in the direction of the line joining them. Even though no conservation laws are broken, such motion was rejected by Bonnor as ``preposterous"\cite{Bonnor89}. However, dismissing the possibility of the existence of an entire class of matter on the basis of the interactions between just two particles may be premature as the dynamics when many particles are present can be quite different, and indeed the results of $N$-body simulations confirm this.

Let us consider then the interactions of Table \ref{tab:table1} in the case when many particles of one or more species are present. For each of the $N$-body simulations carried out, we started with a simple configuration of particles uniformly distributed at random within a sphere of radius 4 and initially at rest, and then let the system evolve, applying the Newtonian gravitational interaction between all particles at each time step using units where $G=|m|=1$ and a time step $\delta t=0.01$. The average radial distance of the particles from the origin is plotted as a function of time, as well as their configuration after a sufficient number of time steps (typically 5,000) have elapsed. We used a simple tree code for efficiency\cite{Pfalzer08} calculating dipole and quadrupole moments of each node for accuracy and using a $\theta$ parameter of 0.5. We carried out each simulation with $N=1920$ particles in keeping with the memory capacity and computational power of the workstation used, and this was sufficient to provide us with a general qualitative understanding of the evolution and dynamics in each case.

Figures \ref{fig:nbodyA} and \ref{fig:nbodyB} show the two types of behaviour that are observed when only one species of particle is present. When there are particles of two different species present, a number of interesting scenarios are possible. Figures \ref{fig:nbodyAB} to \ref{fig:nbodyBC} show the possible types of evolution when two classes of particle are present in equal numbers. The description in the caption of each figure provides more details of the observed dynamics and the expected effect of changing the relative quantities of each particle type, and where appropriate also an indication of how the dynamics is likely to differ in the case that the background spacetime is closed or periodic. The particles of class {\bf A}, {\bf B}, {\bf C} and {\bf D} are coloured black, green, red, and blue, respectively, and lightly coloured particles are more distant than darkly coloured ones.


When more than two classes of particles are present, more complex configurations are possible, and of course the possibilities are endless if more interesting initial conditions are used. A particularly interesting case is the one where particles belonging to all four particle classes are present in equal numbers, and the resulting evolution in this case is shown in Figure \ref{fig:nbodyABCD}.

\section{Discussion and Summary}

We have shown that existence of exotic matter of the types that we have considered are consistent with the laws of classical Newtonian mechanics if the definitions of kinetic energy, momentum and gravitational potential are generalised appropriately.

Although a number of objections have historically been raised against the existence of certain types of exotic matter on account of the unusual and non-intuitive nature of their interactions with ordinary matter, the results of N-body simulations show that in the many-particle case configurations containing such exotic matter can nevertheless evolve to generate the kind of small scale structures and large scale dynamics that are familiar to us from nature.

\vspace{6mm}
\noindent
{\bf References}

\end{document}